\documentclass[final,1p,number,sort&compress]{elsarticle}
\usepackage{amssymb, amsmath}
\usepackage[usenames, dvipsnames]{color}
\newcommand{\be}{\begin{equation}}
\newcommand{\ee}{\end{equation}}
\newcommand{\ber}{\begin{eqnarray}}
\newcommand{\eer}{\end{eqnarray}}
\newcommand{\de}{\end{equation*}}
\newcommand{\cer}{\begin{eqnarray*}}
\newcommand{\der}{\end{eqnarray*}}

\DeclareMathOperator{\re}{Re}

\begin{document}

\begin{frontmatter}
\title{Wigner distribution, nonclassicality and decoherence of generalized
and reciprocal binomial states}
\author[label1]{Anirban Pathak\footnote{anirban.pathak@gmail.com; phone +91 9717066494}}
\address[label1]{Jaypee Institute of Information Technology, A-10, Sector-62,
Noida, UP-201307, India}

\author[label3]{J. Banerji\footnote{jay@prl.res.in}}
\address[label3]{Physical Research Laboratory, Navrangpura, Ahmedabad- 380009, India}
\begin{abstract}
 There are quantum states of light that can be expressed as  finite superpositions of Fock states (FSFS). We demonstrate the nonclassicality of an arbitrary FSFS by means of its phase space distributions such as the Wigner function and the $Q$-function.   The decoherence of the FSFS is studied by considering the time evolution of its Wigner function in amplitude decay  and phase damping channels. As examples, we determine the nonclassicality and decoherence of generalized and reciprocal
binomial states.
\end{abstract}
\end{frontmatter}

\section{Introduction}

In recent years, nonclassical states of light have been recognized as an essential resource in quantum information processing for a variety of applications ranging from the implementation of a class of protocols of discrete \cite{Ekert protocol}
and continuous variable quantum cryptography \cite{CV-qkd-hillery}, to
quantum teleportation \cite{Bennet1993} and dense-coding \cite{densecoding}.
 Additionally, a large number of applications of well known nonclassical
states like squeezed states, entangled states and antibunched states are also reported in recent past. Specifically, squeezed states are
found to be useful for continuous variable quantum cryptography \cite{CV-qkd-hillery} and
teleportation of coherent states \cite{teleportation of coherent state},
entangled states have emerged as one of the most important resources
in quantum information \cite{Nielsen chuang-589} whereas states exhibiting antibunching are useful in building probabilistic single photon sources \cite{antibunching-sps}. All these recently found applications of nonclassical states have added to their importance. However, it is difficult to
maintain nonclassicality (or, quantumness) of a state because of decoherence present in the environment.  Decoherence can not only reduce the amount of nonclassicality, it may also lead to sudden death of a nonclassical character such as entanglement \cite{esd}. The effect of environment that leads to entanglement sudden death and reduction/destruction of other measures of nonclassicality is primarily of two types: amplitude decay and phase damping \cite{esd, esd1}.
These facts have made it very interesting and motivating to systematically investigate the nonclassical properties of quantum states of physical importance and the effect of decoherence (amplitude decay and phase damping channels) on them. The present paper aims to do that for a wide class of quantum states which are finite superpositions of Fock states.

It is well known that Fock states are nonclassical whereas a coherent state, an infinite superposition of Fock states, is `most classical'. Naturally, much interest has been shown in generating and studying the nonclassical properties of states that are {\it finite} superpositions of Fock states (FSFS) (\cite{finitie dimensional states} and references therein). These states can be written in the generic form
\be
|\psi\rangle=\sum_{n=0}^{N}c_{n}|n\rangle,\label{eq:quantum state1}
\ee
where $|n\rangle$ is a Fock state and $N$ is a finite number. Many such states exist in the literature and are named according to the functional form of $c_{n}$. For example, finite dimensional
coherent state \cite{fdc} and most of the intermediate states (e.g.,
binomial state  \cite{D-Stiler}, generalized binomial states
\cite{the:Hong-yi-Fan-generalized-BS,the P Roy0Broy-GBS,the-chenfu-genra;ised-BS},
hypergeometric state \cite{HS}, reciprocal binomial state \cite{RBS-1}  etc.) are of the form (\ref{eq:quantum state1}). Many schemes have been proposed for the experimental generation of FSFS \cite{RBS-2, RBS-3, zou}. Some of the existing proposals are restricted to the realization of specific FSFS \cite{RBS-2, RBS-3} and others are capable to generate an arbitrary FSFS \cite{zou}. For example, Zou, Pahlke and Mathis \cite{zou} provided an interesting scheme for the generation of any arbitrary FSFS using squeezed vacuum states and coherent states as resources.

Nonclassicality of a quantum state can be studied by means of its phase-space distributions such as the Wigner function and the $Q$-function. Negativity in the Wigner function and zeroes of the $Q$-function are signatures of nonclassicality \cite{Agarwal's book} and negative volume in the Wigner function is a quantitative measure of nonclassicality \cite{Kenfack04}. These popular measures of nonclassicality are frequently used to identify and quantify nonclassicality. Specifically,  Wigner functions have often been used to investigate the nonclassicality of a particular FSFS. For example, Wigner functions of binomial states \cite{VIdiella-Baracco}, hypergeometric states (HS) \cite{HS}, excited binomial states \cite{OBES} and truncated coherent states \cite{quantum-scissor} have been studied in detail. In this paper, we  present a compact and generalized framework that can be used to investigate the nonclassicality of any arbitrary FSFS. In order to do so, we derive the Wigner function and the $Q$-function for a FSFS of the form (1).  Interestingly,  the Wigner function reported in the present work is expressed as a finite sum, in contrast to the existing expressions of Wigner functions that are infinite sums and, therefore,  computationally much harder to evaluate. We use our results to explore the nonclassicality of generalized binomial states (GBS) and reciprocal binomial states (RBS). Subsequently, we study the decoherence of these states by considering the time evolution of the Wigner function in amplitude decay and phase damping channels.

The plan of the paper is as follows. In the following subsection,  we briefly introduce generalized and reciprocal binomial states as examples of FSFS. In Section \ref{sec:FSFS},  we derive analytic expressions for different measures of nonclassicality for FSFS. Specifically, we obtain compact expressions for the Wigner function, the $Q$-function, nonclassical volume and time evolution of the Wigner function in an amplitude decay channel and phase damping channel. In Section \ref{sec:gbsandrbs}, we elaborate on the nonclassical character of generalized and reciprocal binomial states. The paper ends with concluding remarks in Section \ref{sec:conclusions}.

\subsection{Generalized  and reciprocal binomial states}

In 1985, Binomial
states were introduced by Stoler \emph{et
al.} \cite{D-Stiler} as states that are  intermediate
between the most nonclassical Fock state $|n\rangle$ and the most
classical coherent state $|\alpha\rangle$. Subsequently several generalizations of binomial states were proposed \cite{the-chenfu-genra;ised-BS,the P Roy0Broy-GBS,the:Hong-yi-Fan-generalized-BS}. The generalized binomial state (GBS)  introduced by Roy and Roy \cite{the P Roy0Broy-GBS} is given as
\be
|N,a,b\rangle=\sum_{n=o}^{N}\sqrt{\omega(n,N,a,b)}|n\rangle
\label{eq:gen-bino1}\ee
 where,
 \be
\omega(n,N,a,b)=\frac{N!}{(a+b+2)_{N}}\frac{(a+1)_{n}(b+1)_{N-n}}
{n!(N-n)!}\label{eq:gen-bino2}\ee
 with $a$, $b >-1$ and $n=0,1,\cdots,N$. Here, $(x)_{n}$ denotes the usual Pochhammer
symbol:
\be
\begin{array}{lcr}
(x)_{0}=1 & \,\,\,\,\,\,\, & (x)_{n}=x(x+1)\cdots(x+n-1).\end{array}\label{eq:gen-bino3}\ee
This form of GBS is of particular interest as it reduces to
the vacuum state, Fock state, coherent state, binomial state and negative binomial state in different limits of $a$, $b$ and $N$.

Another interesting state of the form (1)
is the reciprocal binomial state (RBS). It was introduced by Barnett and
Pegg \cite{RBS-1} in 1996 as a reference quantum state that can be
used to measure the quantum optical phase probability distribution
using projection synthesis method. Subsequently a number of experimental schemes
for the generation of RBS was proposed \cite{RBS-2, RBS-3}.  Moussa and Baseia first proposed
a scheme for the experimental generation of RBS using $N$ identical two-level circular Rydberg atoms, a Ramsey zone, a high-$Q$ cavity and a state selective detector \cite{RBS-2}. Later on Valverde et al. had shown that it is possible to generate RBS using an
array of beam-splitters \cite{RBS-3}. RBS can
also be viewed as a generalization of binomial state (where the coefficients
are inverse of that in the binomial state) and is defined as \cite{RBS-1,RBS-2}
\be
|\phi\rangle=\frac{1}{\cal N}\sum_{k=0}^N \left(^{N}C_{k}\right)^{-1/2}
e^{ik(\phi-\pi/2)}|k\rangle\label{eq:rbs1}
\ee
 where $\cal N$ is a normalization constant: ${\cal N}=\sqrt{\sum_{k=0}^{N}\left(^{N}C_{k}\right)^{-1}}$.

\section{Nonclassicality of a Finite Superposition of Fock States (FSFS) \label{sec:FSFS}}
From the negativity of the Wigner function and zeroes of the $Q$-function we can obtain  signatures
of nonclassicality in a FSFS.
\subsection{Wigner function}
In \cite{series-wigner}, the authors commented, ``The
Wigner function is usually expressed in an integral form which is
not always easy to compute.'' For a state with density matrix $\rho$, they provided
(cf. Eqn. 16 of \cite{series-wigner} ) the following analytic expression
of the Wigner function: \begin{equation}
W(\alpha)=\frac{2}{\pi}\sum_{k=0}^{\infty}(-1)^{k}\langle\alpha,k|\rho|\alpha,k\rangle,\label{eq:series wigner}\end{equation}
where  $|\alpha,k\rangle=D(\alpha)|k\rangle$
are the displaced number states and $D(\alpha)=\exp(\alpha a^\dagger -{\alpha^*} a)$ is the displacement operator.

 Till now, the  above mentioned infinite-series expansion of Wigner function has been used even for FSFS such as the hypergeometric state \cite{HS} and the excited binomial state \cite{OBES}. Since $|\psi\rangle$ in (\ref{eq:quantum state1})
is a pure state, we have $\rho=|\psi\rangle\langle\psi|$ and consequently,
\be
W(\alpha)  = \frac{2}{\pi}\sum_{k=0}^{\infty}(-1)^{k}|\sum_{n}c_{n}\chi_{kn}(\alpha)|^{2},\ee
where $\chi_{kn}(\alpha)$ has the following expression
(cf. Eqn. 4.6 of \cite{HS})
\be
\chi_{nk}(\alpha)=\left\{ \begin{array}{c}
\sqrt{k!/n!}\,\exp(-|\alpha|^{2}/2)\alpha^{n-k}L_{k}^{n-k}\left(|\alpha|^{2}\right)\,{\rm if\:}n\geq k\\\\
\sqrt{n!/k!}\,\exp(-|\alpha|^{2}/2)\left(\alpha^{*}\right)^{k-n}L_{n}^{k-n}\left(|\alpha|^{2}\right)\,\,{\rm if\:}n\leq k.\end{array}\right.\ee
In what follows, we show that for a FSFS of the form (1), the above prescription is unnecessary as the relevant integration can be done analytically and the Wigner function can be expressed as a {\it finite} sum.

In the coherent state representation, the Wigner function of a state $\vert \psi\rangle$ is given by
\be
W\left(\gamma,\gamma^* \right ) = \frac{2}{\pi^{2}} \exp(2\vert\gamma\vert^{2}) \int \langle-\lambda\vert\rho\vert\lambda\rangle \exp\{-2\left(\lambda\gamma^* -\lambda^* \gamma\right)\} {\mathrm{d}^{2}\lambda}.
\ee
 For a FSFS of the form (1), we use the Fock state decomposition of coherent states $\vert \lambda\rangle$ to obtain
\be
W\left(\gamma,\gamma^* \right ) =\sum_{n=0}^N \sum_{m=0}^N c_n c^*_m X_{nm},
\ee
where
\ber
\lefteqn{X_{nm} =  \frac{2 (-1)^n}{\pi \sqrt{n!m!}} \exp(2\vert\gamma\vert^{2})} \nonumber\\
&& {}\times\int \frac{{\mathrm{d}^{2}\lambda}}{\pi} {\lambda^*}^n {\lambda}^m \exp[-\vert\lambda\vert^2 -2(\gamma^* \lambda -\gamma \lambda^*)].
\eer
The integration over $\lambda$ is carried out by using the formula \cite{magnus}
\be
\int \frac{{\mathrm{d}^{2}z}}{\pi} {z^*}^n {z}^m \exp(\xi\vert z\vert^2 +\sigma z +\eta z^*)=\exp(-\sigma \eta/\xi)\sum_{j=0}^{\min (m,n)}\frac{m!n! \sigma^{n-j}\eta^{m-j}}{j!(n-j)! (m-j)! (-\xi)^{m+n-j+1}}.
\ee
We obtain
\ber
X_{nm}&=& \frac{2 (-1)^n}{\pi}\sqrt{n!m!}\exp(-2\vert \gamma\vert^2) (2\gamma)^{m-n}\nonumber\\
&& \times \sum_{j=0}^n \frac{(-2\gamma^*)^{n-j}(2\gamma)^{n-j}}{j!(n-j)! (m-j)!}\\
&=& \frac{2 (-1)^n}{\pi}\sqrt{\frac{n!}{m!}}\exp(-2\vert \gamma\vert^2) (2\gamma)^{m-n}L_n^{m-n}(4\vert \gamma\vert^2)\qquad n<m.
\eer
For $n<m$, it is easy to show that $X_{mn}=X_{nm}^*$. Thus, $W(\gamma,\gamma^*)$ can be written as
\be
W(\gamma,\gamma^*)=\sum_{n=0}^N \vert c_n\vert^2 X_{nn}+ 2\re \sum_{m=1}^N\sum_{n=0}^{m-1}c_n c^*_m X_{nm}.
\label{eq:wig1}\ee
\subsection{$Q$-function}
The $Q$-function for a state $\vert \psi\rangle$ is given as $Q (\gamma)=\vert \langle \gamma \vert\psi\rangle\vert^2$,  where $\vert\gamma\rangle$ is a coherent state. For a state of the form (1), we immediately get
\be
Q(\gamma)=\frac{\exp(-\vert\gamma\vert^2)}{\pi}\left\vert\sum_{n=0}^N c_n \frac{{\gamma^*}^n}{\sqrt{n!}} \right\vert^2.
\label{eq:q1}\ee

\subsection{Nonclassical volume}
Negativity of the Wigner function and zeroes of the $Q$-function provide us the signatures of
nonclassicality, but we cannot obtain any idea about the amount of nonclassicality through them.
There exist several quantitative measures
of nonclassicality. Here we may use the most relevant quantitative
measure of nonclassicality which is known as the nonclassical volume.
The nonclassical volume as a measure of quantumness was first introduced
by Kenfack and Zyczkowski \cite{Kenfack04} in 2004. In this particular {\it quantitative}
measure, the volume of the negative part of the Wigner function is
considered as the measure of nonclassicality. The negative
volume associated with a quantum state $|\psi\rangle$ is

\be
\delta=\frac{1}{2}\left(\int\int \vert W(\gamma,\gamma^*)\vert {\mathrm{d}^{2}\gamma}-1\right).
\ee
A non-zero value of $\delta$ indicates a nonclassical
state. As we have compact expression of the Wigner function of FSFS, we can
use it to obtain $\delta$ for various choices of parameters
and investigate how nonclassical volume (or the amount of quantumness)
varies with the change of a particular parameter.

\subsection{Decoherence}
The effect of decoherence on the Wigner function and hence the nonclassicality of a state can be studied by using the approach of Biswas and Agarwal \cite{Ashoka-and-Agarwal}. Decoherence can be due to loss of photons to the reservoir (amplitude decay) or phase damping.

In the amplitude decay model, the evolution of the Wigner function is given by
\be
W(\gamma,\gamma^*,t)=\frac{2}{1-\beta^{2}}\int \frac{{\mathrm{d}^{2}\gamma}}{\pi} W(\gamma_0,\gamma_0^*)\exp\left(-\frac{2\vert \gamma-\beta \gamma_0\vert^2}{1-\beta^{2}}\right),\qquad \beta=\exp(-\kappa t)
\label{eq:decay1}.\ee
Substituting from (10) and (13) and using the formula (12), one can write
\be
W(\gamma,\gamma^*,t)=\sum_{n=0}^N \vert c_n\vert^2 Y_{nn}+ 2\re \sum_{m=1}^N\sum_{n=0}^{m-1}c_n c^*_m Y_{nm}
\ee
where,
\ber
\lefteqn{Y_{nm} =  \frac{2 (-1)^n}{\pi} \sqrt{n!m!} \exp(-2\vert\gamma\vert^{2}) (2\beta\gamma)^{m-n}}\nonumber\\
&& {}\times\sum_{j=0}^n \frac{(-2)^j (1-\beta^{2})^j L_j^{m-n}\left(-\frac{2\beta^{2}\vert \gamma\vert^2}{1-\beta^{2}}\right)}{(n-j)!(m-n+j)!}\qquad n\leq m.\label{ynm}
\eer
It is easy to see that at $t=0$ ($\beta=1$), $Y_{nm}\to X_{nm}$, as expected.

We can use the formalism presented here to study the nonclassical properties of any FSFS. In the next section, we use the expressions obtained here to explore the nonclassicality and decoherence of GBS and RBS for various choices of the parameters defining these states.

\section{Nonclassicality and decoherence of GBS and RBS \label{sec:gbsandrbs}}
\begin{figure}[htbp]
\centerline{\includegraphics[width=14cm]{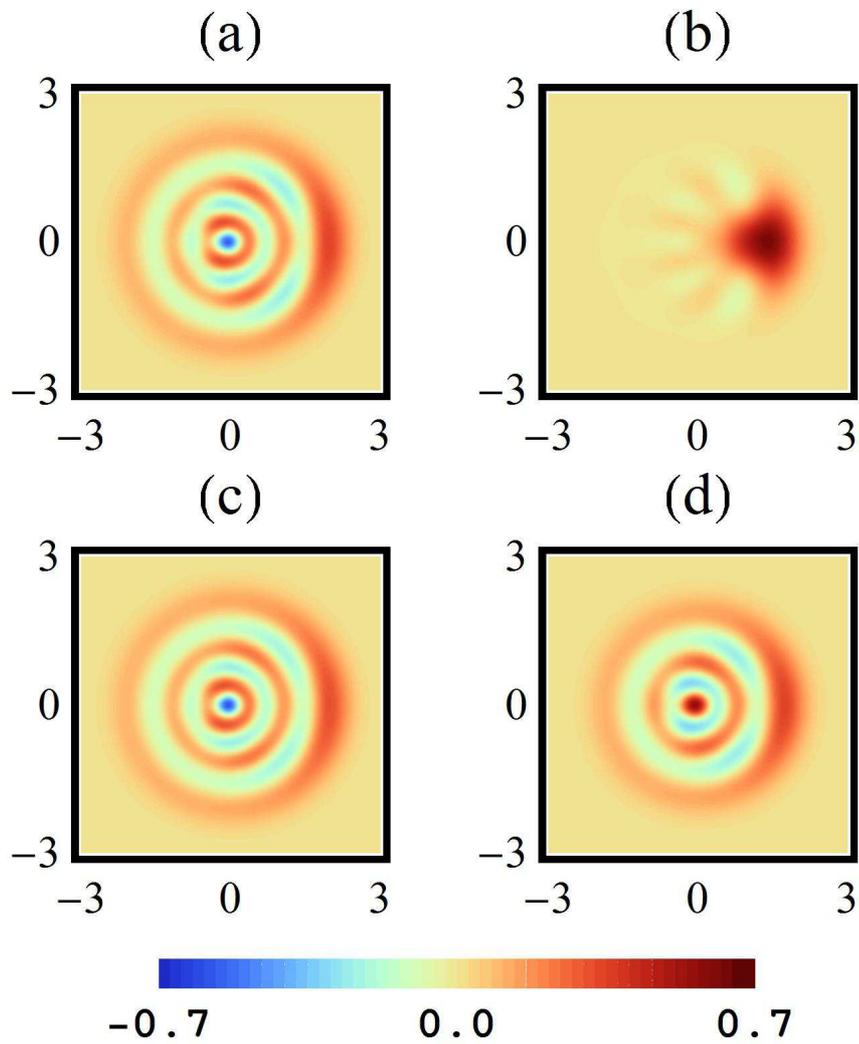}}
\caption{\label{fig:Wigner-GBS}(Color online) Contour plot of the Wigner function of GBS for
(a) $N=5,\, a=0.9,\, b=-0.9,$ (b) $N=5,\, a=0.9,\, b=2,$  (c) $N=5,\, a=2,\, b=-0.9,$
(d) $N=4,\, a=0.9,\, b=-0.9$. Nonclassicality decreases with increase in $b$.}
\end{figure}

\begin{figure}[htbp]
\centerline{\includegraphics[width=14cm]{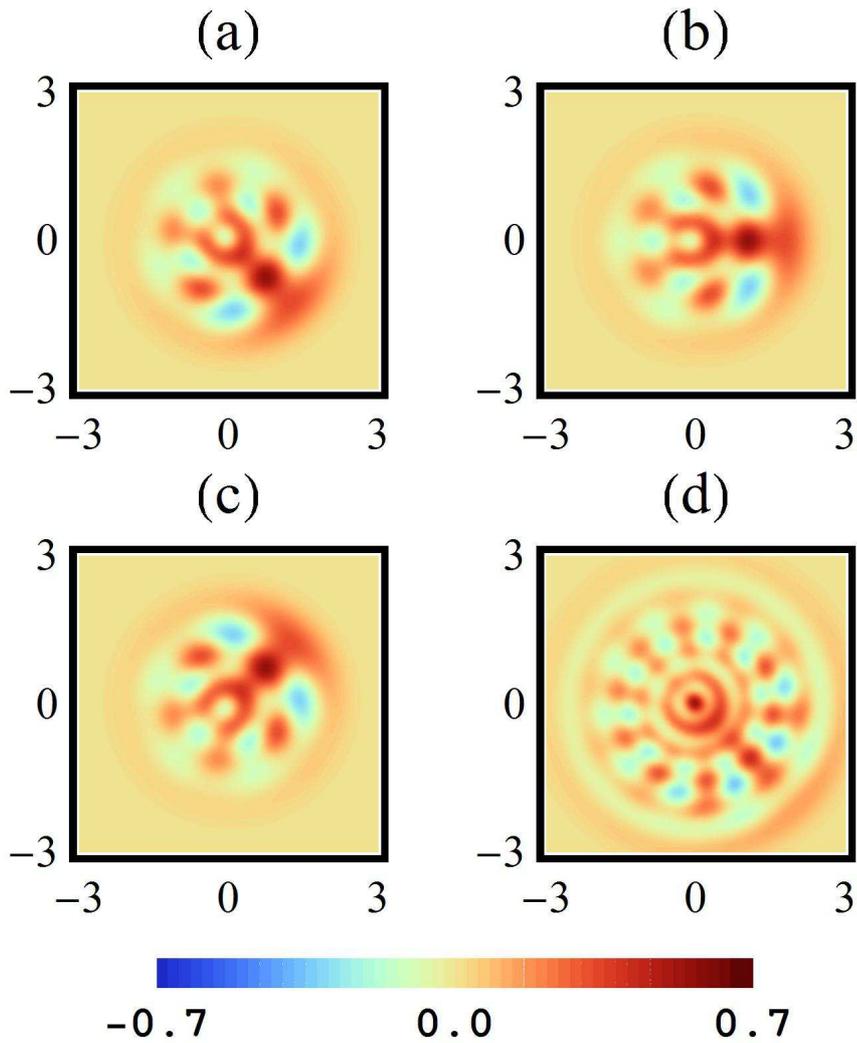}}
\caption{\label{fig:Wigner-RBS}(Color online) Contour plot of the Wigner function of RBS for
(a) $\phi=\pi/4,\, N=5$, (b) $\phi=\pi/2,\, N=5$,
(c) $\phi=3\pi/4,\, N=5$ and (d) $\phi=\pi/4,\, N=10$.
The Wigner function mostly rotates with change in $\phi$. However,
nonclassicality increases with increase in $N$.}
\end{figure}

The Wigner functions
for the GBS and the RBS are obtained by using (\ref{eq:wig1}) and shown in Figs. \ref{fig:Wigner-GBS} and  \ref{fig:Wigner-RBS} respectively. Clearly there
exist negative regions of Wigner function in all the figures and these negative regions
illustrate that these two FSFSs (i.e., GBS and RBS) are nonclassical for various choices of parameters.

At the center of the phase space ($\gamma=\gamma^*=0$), the Wigner function has the value
\be
W(0,0)=\frac{2}{\pi}\sum_{n=0}^N (-1)^n \vert c_n\vert^2.\label{w00}
\ee
Clearly, the overall sign of $W(0,0)$ depends on how $|c_n|^2$ varies with $n$. As an example, let us consider a GBS with $a=0.9$ and $b=-0.9$ (see Fig. \ref{fig:Wigner-GBS} (a) and (d)). For $N=5$,
\be
\begin{array}{lll}
\vert c_0\vert^2= 0.00407779, & \vert c_1\vert^2=0.00944854, & \vert c_2\vert^2=0.0176779,\\
\vert c_3\vert^2=0.0328304, & \vert c_4\vert^2=0.0731223, & \vert c_5\vert^2= 0.862843.
\end{array}
\ee
whereas for $N=4$,
\be
\begin{array}{lll}
\vert c_0\vert^2=0.0059675, & \vert c_1\vert^2=0.01463, & \vert c_2\vert^2=0.030305,\\ \vert c_3\vert^2=0.07163, & \vert c_4\vert^2= 0.877468.
\end{array}
\ee
Thus, for $a=0.9$ and $b=-0.9$ and a given value of $N$, $|c_{n+1}|^2>|c_n|^2$. Furthermore, $|c_N|^2$ is much greater than $|c_n|^2$ ($n\neq N$). Thus $W(0,0)$ is negative if $N$ is odd (as in Fig. 1 a) and positive if $N$ is even (as in Fig. 1 d). For other values of $a$ and $b$, $|c_n|^2$ will have a different distribution with respect to $n$ and consequently, the sign of $W(0,0)$ will change accordingly.

For a RBS with a given value of $N$, $|c_n|^2$ is a function of $n$ only. Using the relation ${}^{N}C_{N-k}={}^{N}C_{k}$, it is easy to show that $W(0,0)$ is zero for odd $N$ (as in Fig. 2 a) and greater than zero for even $N$ (as in Fig. 2 d).

\begin{figure}[htbp]
\centerline{\includegraphics[width=8cm]{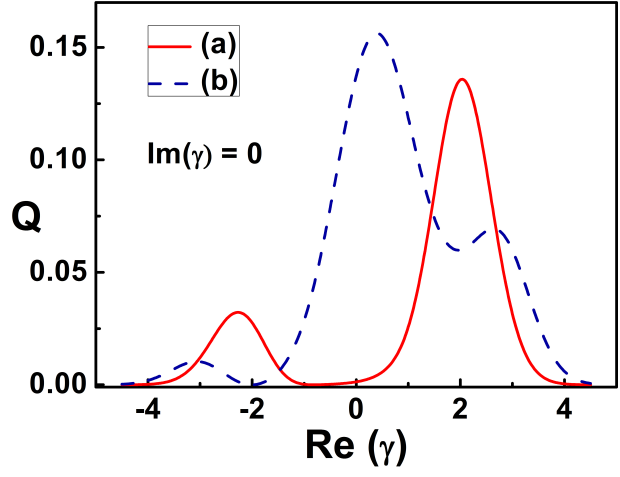}}
\caption{\label{fig:Q-function}(Color online)  The $Q$-function of GBS (solid line) for $N=5$, $a=0.9$, $b=-0.9$ and RBS (dashed
line) for $N=9$ and $\phi=\pi/2$. We have set $\rm{Im}(\gamma)=0$. Clearly, in both  cases, the $Q$-function has zeroes
and consequently it depicts nonclassicality.}
\end{figure}

Additionally, as shown in Fig. \ref{fig:Q-function}, holes in the $Q$-functions can be detected by using (\ref{eq:q1}). These holes are signatures of nonclassicality.

\begin{figure}[htbp]
\centerline{\includegraphics[width=8cm]{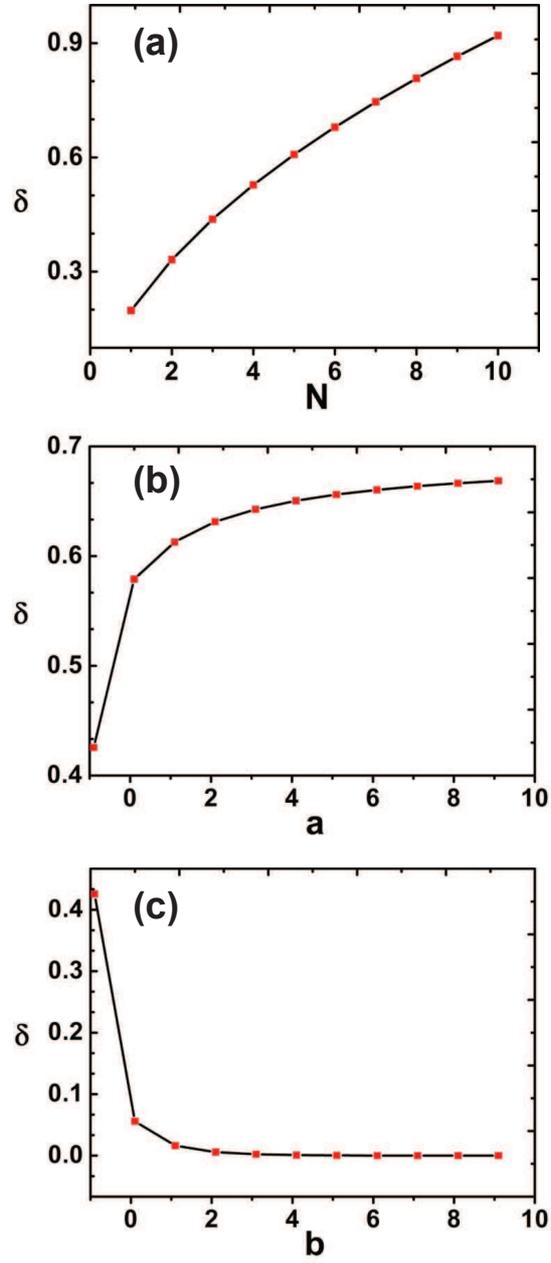}}
\caption{\label{fig:nonclassicalvolume-GBS}(Color online)   The variation of nonclassical volume
 $\delta$ for GBS (a) with $N$, for $a=0.9, b=-0.9$,  (b)  with $a$, for $N=5$, $b=-0.9$, (c) with $b$, for $N=5, a=-0.9$.}
\end{figure}

\begin{figure}[htbp]
\centerline{\includegraphics[width=8cm]{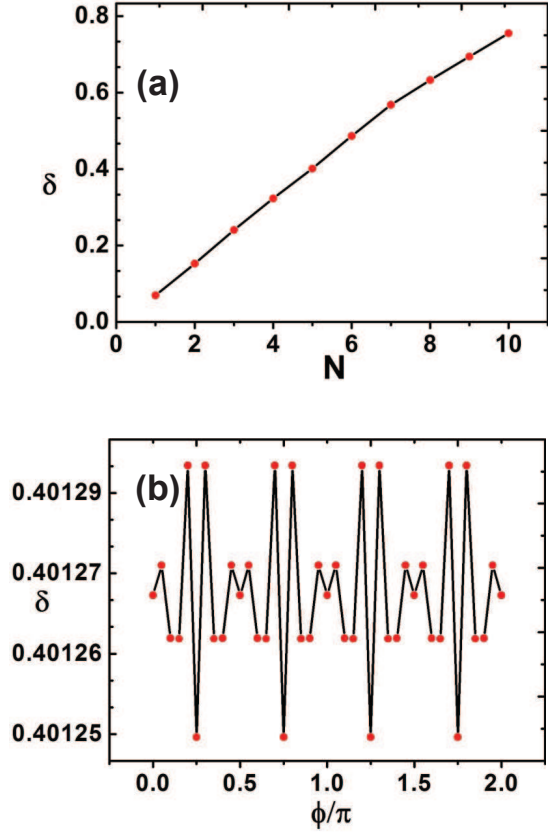}}
\caption{\label{fig:nonclassicalvolume-RBS}(Color online) The variation of nonclassical volume
 $\delta$ for RBS (a) with $N$, for $\phi=\pi/2$,  (b)  with $\phi$, for $N=5$.}
\end{figure}

 In  Fig. \ref{fig:nonclassicalvolume-GBS} we have shown the variation of nonclassical volume $\delta$ for GBS
with $N$, $a$ and $b$ and have observed that the amount of nonclassicality
increases with the increase of $N$ and $a$ but it decreases with the increase
of $b$. Similarly, we illustrate the variation of $\delta$
for RBS with $N$ and $\phi$ in Fig. \ref{fig:nonclassicalvolume-RBS}
and found that the amount of nonclassicality increases with $N$ and has a slight periodic variation in $\phi$ with a periodicity of $ \pi/2$.

\subsection{Effect of decoherence}

\begin{figure}[htbp]
\centerline{\includegraphics[width=8cm]{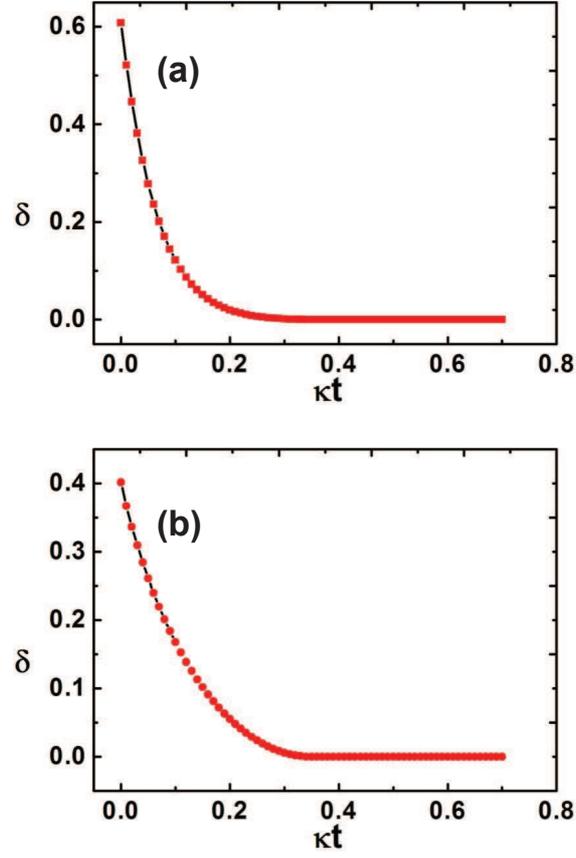}}
\caption{\label{fig:decay}(Color online)  The decay of nonclassical volume
 $\delta$ with time in the amplitude decay channel for (a)  GBS with $N=5$, $a=0.9$, $b=-0.9$, and (b)  RBS with $N=5$, $\phi=\pi/2$.}
\end{figure}

In this section we have studied the effect of decoherence on the nonclassicality
using the approach of Biswas and Agarwal \cite{Ashoka-and-Agarwal}. Specifically, we  have used (\ref{eq:decay1}) to investigate the
effect of decoherence (amplitude decay channel) on the Wigner functions of GBS and RBS.  In Fig.  \ref{fig:decay} the variation of $\delta$ with $\kappa t$ is shown for both GBS and RBS. Clearly, in both cases,  nonclassical volume decays rapidly with rescalled time ($\kappa t$)  leading to a loss of nonclassicality.

In the presence of decoherence due to amplitude decay,  the Wigner function at the origin of phase space has the expression
\be
W(0,0,t)=\frac{2}{\pi}\sum_{n=0}^N (1-2\beta^2)^n \vert c_n\vert^2.\label{w00t}
\ee
For $t=0$, the above expression reduces to (\ref{w00}). Note that, for $\kappa t>\frac{1}{2}\ln {2}$, $W(0,0,t)$ is positive for all parameter values that define $c_n$.
Finally, we consider the limit $t\to \infty$. From (\ref{eq:decay1}) or (\ref{ynm}), we observe that $W(\gamma,\gamma^*,\infty)=2/\pi \exp(-2|\gamma|^2)$ which is the Wigner function for the vacuum state, i.e., all the photons are lost to the reservoir.

\begin{figure}[htbp]
\centerline{\includegraphics[width=8cm]{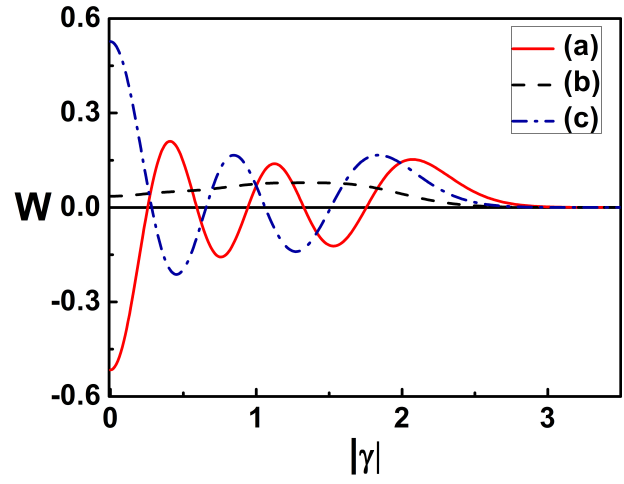}}
\caption{\label{wgbs}(Color online) The Wigner function at $t=\infty$ under decoherence due to phase damping is a function of $|\gamma|$ only. Shown here are the plots for a GBS with (a) $N=5$, $a=0.9$ $b=-0.9$, (b) $N=5$, $a=0.9$ $b=2.0$, and (c) $N=4$, $a=0.9$ $b=-0.9$.}
\end{figure}

\begin{figure}[htbp]
\centerline{\includegraphics[width=8cm]{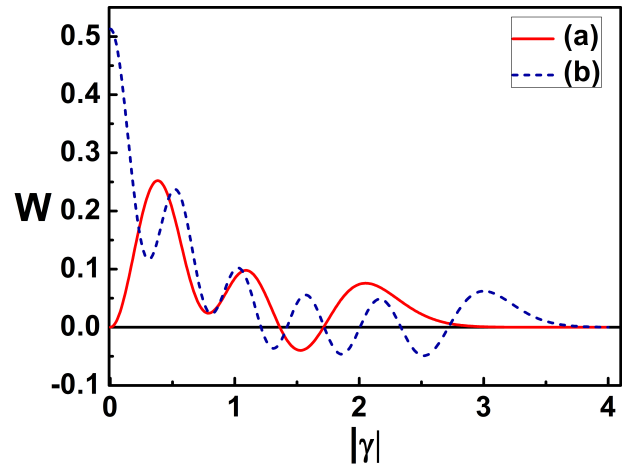}}
\caption{\label{wrbs}(Color online)  As in Fig. 7, but for a RBS with $\phi=\pi/4$ and (a) $N=5$, (b) $N=10$.}
\end{figure}

Apart from amplitude damping channel, occurrence of phase damping channel is also very common in nature and they are known to play a crucial role in entanglement sudden death \cite{esd, esd1, esd2}.  We  note briefly that in the case of decoherence due to phase damping \cite{Ashoka-and-Agarwal}, $W(\gamma,\gamma^*,\infty)=\sum_{n=0}^N |c_n|^2 X_{nn}$. This is circularly symmetric (function of $|\gamma|$ only) in phase space. In Figs. \ref{wgbs} and \ref{wrbs}, we plot $W(\gamma,,\gamma^*,\infty)$ for some examples of GBS and RBS. In both cases, even in the limit $t\to \infty$, the states remain nonclassical as their Wigner functions exhibit negativity.

\section{Conclusions\label{sec:conclusions}}

We have derived compact analytic expressions for the Wigner function and the $Q$-function of an arbitrary FSFS to study its nonclassical character and decoherence.  As examples,  we have applied our theory to the generalized and reciprocal binomial states.
It is a straightforward exercise to similarly study the nonclassicality and decoherence of other types of FSFS such as
the  hypergeometric state, the binomial
state, and the finite dimensional coherent state.

We end by noting that several schemes for the direct measurement of Wigner function have been proposed in recent years \cite{direct wigner measurement1, direct wigner measurement2, direct wigner measurement3, direct wigner measurement4}.
Applicability of most of these schemes are limited to specific quantum systems. However, keeping in mind the recent progress in the direct measurement of Wigner function, we hope that it will be possible to experimentally
verify the results of present theoretical investigation. Alternatively,  one can reconstruct the Wigner function from a directly measurable quantity, the optical tomogram.  The optical tomogram $w_{|\psi\rangle}(X,\theta)$ of a quantum state
$|\psi\rangle$ may be visualized as the marginal distribution of
the quadrature component $X$ of the electric field strength, with
$X$ being rotated by angle $\theta$ in the quadrature phase space
\cite{Manko 1}. Once a tomogram is obtained, we may use the optical
homodyne tomographic technique described in \cite{optical-homodyne-tomography}
to obtain the Wigner function and thus experimentally verify the
Wigner functions reported in the present paper. Interestingly, Filippov
and Man'ko \cite{Manko 1} have recently reported a closed form analytic
expression for the optical tomogram of (\ref{eq:quantum state1})
as
\ber
\lefteqn{w_{|\psi\rangle}(X,\theta)= \frac{e^{-X^{2}}}{\sqrt{\pi}}\left[\sum_{n=0}^{N}\frac{|c_{n}|^{2}}{2^{n}n!}H_{n}^{2}(X)\right.}\nonumber\\
&&{}\left.+\sum_{n<k}\frac{|c_{n}||c_{k}|\cos\left(\left(n-k\right)\theta-\left(\phi_{n}-\phi_{k}\right)\right)}{\sqrt{2^{n+k-2}n!k!}}H_{n}(X)H_{k}(X)\right]\label{eq:optical tomogram}
\eer
where $c_{j}=|c_{j}|e^{i\phi_{j}}$ and $H_{j}$ is the Hermite polynomial
of degree $j$. They have used the expression
to construct optical tomograms of some simple quantum states, which are superposition of vacuum and low-photon Fock states (e.g., $|0\rangle+i|1\rangle$, $|0\rangle+|2\rangle$
etc.). Likewise, one can use (\ref{eq:optical tomogram}) to obtain optical tomograms
of the GBS and the RBS.
 Thus the results
reported in the present paper based on the  Wigner function can be experimentally
verified by using optical homodyne tomography and the expected experimental
tomograms can be theoretically obtained from (\ref{eq:optical tomogram}).

\textbf{Acknowledgments}: A.P. thanks Department of Science and Technology (DST), India for support provided through the DST project No. SR/S2/LOP-0012/2010.

\clearpage
\section*{List of Figure Captions}
Fig.1. (Color online) Contour plot of the Wigner function of GBS for
(a) $N=5,\, a=0.9,\, b=-0.9,$ (b) $N=5,\, a=0.9,\, b=2,$  (c) $N=5,\, a=2,\, b=-0.9,$
(d) $N=4,\, a=0.9,\, b=-0.9$. Nonclassicality decreases with increase in $b$.
\par
\noindent Fig. 2. (Color online) Contour plot of the Wigner function of RBS for
(a) $\phi=\pi/4,\, N=5$, (b) $\phi=\pi/2,\, N=5$,
(c) $\phi=3\pi/4,\, N=5$ and (d) $\phi=\pi/4,\, N=10$.
The Wigner function mostly rotates with change in $\phi$. However,
nonclassicality increases with increase in $N$.
\par
\noindent Fig. 3. (Color online)  The $Q$-function of GBS (solid line) for $N=5$, $a=0.9$, $b=-0.9$ and RBS (dashed
line) for $N=9$ and $\phi=\pi/2$. We have set $\rm{Im}(\gamma)=0$. Clearly, in both  cases, the $Q$-function has zeroes
and consequently it depicts nonclassicality.
\par
\noindent Fig. 4. (Color online)   The variation of nonclassical volume
 $\delta$ for GBS (a) with $N$, for $a=0.9, b=-0.9$,  (b)  with $a$, for $N=5$, $b=-0.9$, (c) with $b$, for $N=5, a=-0.9$.
\par
\noindent Fig. 5. (Color online) The variation of nonclassical volume
 $\delta$ for RBS (a) with $N$, for $\phi=\pi/2$,  (b)  with $\phi$, for $N=5$.
\par
\noindent Fig. 6. (Color online)  The decay of nonclassical volume
 $\delta$ with time in the amplitude decay channel for (a)  GBS with $N=5$, $a=0.9$, $b=-0.9$, and (b)  RBS with $N=5$, $\phi=\pi/2$.
\par
\noindent Fig. 7. (Color online) The Wigner function at $t=\infty$ under decoherence due to phase damping is a function of $|\gamma|$ only. Shown here are the plots for a GBS with (a) $N=5$, $a=0.9$ $b=-0.9$, (b) $N=5$, $a=0.9$ $b=2.0$, and (c) $N=4$, $a=0.9$ $b=-0.9$.
\par
\noindent Fig. 8. (Color online)  As in Fig. 7, but for a RBS with $\phi=\pi/4$ and (a) $N=5$, (b) $N=10$.

\clearpage
\clearpage
\clearpage
\clearpage
\clearpage
\clearpage
\clearpage
\clearpage
\end{document}